\documentclass[12pt, oneside]{article}

\usepackage{epsfig}
\usepackage{latexsym}
\usepackage{amssymb}
\usepackage{amsmath}
\usepackage{graphicx}
\usepackage{subfigure} 
\usepackage{amsfonts}
\usepackage{array}

\usepackage{fancyhdr}
\fancypagestyle{plain}{
 \fancyhead{} 
 }

\makeatletter
\def\cleardoublepage{\clearpage\if@twoside \ifodd\c@page\else%
	     \hbox{}%
	 \thispagestyle{empty}
	 \newpage%
	 \if@twocolumn\hbox{}\newpage\fi\fi\fi}
\makeatother

\numberwithin{equation}{section}

\newcommand{\be}{\begin{equation}}
\newcommand{\ee}{\end{equation}}
\newcommand{\bea}{\begin{eqnarray}}
\newcommand{\eea}{\end{eqnarray}}

\def\gsim{\raise0.3ex\hbox{$>$\kern-0.75em\raise-1.1ex\hbox{$\sim$}}}
\def\lsim{\raise0.3ex\hbox{$<$\kern-0.75em\raise-1.1ex\hbox{$\sim$}}}

\begin{document}

\title{\textbf{Higher Dimensional Choptuik Scaling in Painleve Gullstrand Coordinates}}

\author{{\bf Tim Taves$^*$ and Gabor Kunstatter$^\dagger$}\\[10pt]
{\small $^*$ Department of Physics and Astronomy}\\
{\small University of Manitoba, R3T 2N2 Canada}\\ 
{\small and Winnipeg Institute for Theoretical Physics,}\\
{\small   Winnipeg, Manitoba, Canada}\\[5pt] 
{\small $^\dagger$ Physics Department, University of Winnipeg,}\\
{\small Winnipeg, MB, R3B 2E9, Canada and} \\
{\small Winnipeg Institute for Theoretical Physics,}\\
{\small  Winnipeg, Manitoba, Canada}}

\date{\today}

\maketitle

\begin{abstract}
We investigate Choptuik scaling in the spherically symmetric collapse of a massless scalar field in higher dimensions using Painleve-Gullstrand (PG) coordinates. Our analysis gives reliable results for the critical exponent and period echoing up to seven dimensions and confirms the presence in higher dimensions of the cusps in the periodic scaling relationship recently observed in four dimensional collapse. In addition, our value for the critical exponent in seven dimensions is consistent with that obtained by Bland {\it et al}, who argued that the critical exponent increases monotonically with dimension to an asymptotic value of 1/2.


\end{abstract}

\clearpage

\section{Introduction}

The formation of a singularity at the center of black holes signals the breakdown of classical general relativity.  It is speculated that a quantum theory of gravity would resolve the singularity problem.  Some popular theories of quantum gravity, such as string theory, depend upon the existence of more than four dimensions. It has even been suggested that microscopic higher dimensional black holes could be produced in the LHC, and experimental bounds on their production have appeared in the literature \cite{bhlhc}.  It is therefore important to understand the details of black hole formation in higher dimensions.

In the early 90's Matthew Choptuik \cite{MWC93} numerically simulated the collapse of a spherically symmetric massless scalar field and found a simple universal scaling relationship between physical observables describing the black hole on formation and the parameters describing the initial scalar field distribution. Since then, this universal scaling has been confirmed in a large number of different settings. (For a review of Choptuik scaling see \cite{CG03}.) To be more precise,  as a single parameter, $A$, for any family of initial data, is varied the initial data is separated into two classes. For $A$ less than some critical value $A_*$, the matter bounces from the origin and ultimately dissipates. For $A>A_*$, an event horizon forms and the end state is a black hole. As $A \to A_*$ from above, the final spacetime approaches arbitrarily closely the critical black hole solution which in effect describes a zero mass black hole that exhibits a naked singularity at the center. Near criticality, any geometrical quantity, $X$, characterizing the evolution of the black hole at the time of formation, obeys a scaling law of the form:
\be
ln(X) = \gamma ln|A-A_*|+f(ln|A-A_*|),
\label{Chop}
\ee
where $\gamma$ is the critical exponent and $f$ is a periodic function with period, $T$. $\gamma$ and $T$ are universal in the sense that they are independent of the quantity being measured and also of the type of initial data. They do depend on the type of matter that is collapsing, however. The specific form of the oscillatiing function, $f$, is known not to be universal. The most common quantities used to test this relationship are the mass of the final black hole (or equivalently the radius of the event horizon) and the maximum value of the curvature in subcritical collapse.
 Previous calculations for the collapse of a spherically symmetric massless scalar field in four spacetime dimensions in Schwarzschild, double null and Painleve Gullstrand (PG) coordinates have all given $\gamma \approx 0.37$ and $T \approx 4.6$ \cite{MWC93,CG97,SHTP97,JZ09,JB05,DG95,RSH96}.  
 
In Schwarzschild and double null coordinates the form of the periodic function, $f$, is well approximated by a small amplitude sine function.  By contrast, in PG coordinates the periodic function in four dimensions showed distinctive cusps \cite{JZ09}.  This difference in the form of $f$ in PG coordinates is perhaps surprising, but not inconsistent. The PG calculation did not measure the final black hole mass after all matter had fallen through the horizon, which is independent of slicing and would exhibit the same behaviour in all coordinates. Instead, what was plotted was the radius of the apparent horizon on formation, a quantity that does depend on the slicing. It is nonetheless a geometrical quantity that exhibits Choptuik scaling, as shown in \cite{JZ09}.  What is less clear is whether the cusp-like features of the scaling function is a peculiarity of PG coordinates in 4 dimensions or generic in some sense.

Most previous calculations of Choptuik scaling have involved four dimensional black holes. The first higher dimensional analysis was done by Garfinkle and Duncan \cite{CCD99}. More recently, a program was initiated \cite{BHKVO02} whose purpose was to calculate the critical exponent and echoing period for spherical collapse in arbitrary spacetime dimensions using the so-called dilaton gravity formalism. Preliminary results were obtained in 5 and 6 spacetime dimensions. Subsequently, accurate results were obtained up to D=14 \cite{JB05,JBthesis}. These results provide strong evidence that the critical exponent was a monotonic function of D that converged asymptotically to 1/2. Also around that time, Oren and Sorkin\cite{ES05} produced results for the critical exponent up to D=11 which they used to speculate that the critical exponent was not a monotonic function, but instead achieved a maximum near D=10. Moreover, it was argued on the basis of the behaviour of black string solutions that a critical dimension near D=10 might not be unexpected. It should be noted that both sets of calculations were done using double null coordinates, using different parametrizations of the fields.  

The higher dimensional calculations referred to above were done in double null coordinates, whereas the purpose of the present paper is to look at spherically symmetric scalar field collapse in higher dimensions using Painleve-Gullstrand (P-G) coordinates. It is perhaps worth highlighting the relative merits of the two different sets of coordinates. The main advantage of double null coordinates for studying critical behaviour has to do with the fine spatial resolution that is possible due to the convergence of outgoing null rays near horizon formation. However, this same convergence makes it difficult, if not impossible  to allow numerical simulations to run all the way up to horizon formation (see numerics section of \cite{DG95}).  For this reason the mass/horizon radius of the black hole in null coordinates must be approximated by choosing in advance how close one wishes to get to horizon formation.  This problem is also encountered in Schwarzschild coordinates. PG coordinates provide spatial slices that are regular across future horizons, so that the code can be run up to and even beyond horizon formation. Away from criticality, this allows one to map out the structure of the trapping horizon as done in \cite{JZ09}. Near criticality, we can precisely (to numerical accuracy) determine the radius of the horizon on formation. Note that this quantity is different  from the mass of the final black hole. It marks the location that the P-G slicing intersects the trapping horizon and therefore verifies the scaling relationship in a different geometrical quantity therbye providing independent measures of the critical exponent and echoing scale. This also explains why a different oscillatory function is possible. The main disadvantage of P-G coordinates is that one does not get an automatic improvement in the spatial resolution from the convergence of null rays. Moreover, as discussed in \cite{JZ09} the time steps are limited by the null cone structure.

As with previous calculations, the resulting numerics get more difficult as the number of spacetime dimensions is increased. We obtain reliable results  in 4, 5, 6 and 7 dimensions for the scaling law obeyed by the areal radius of the apparent horizon on formation.  We investigate the periodic function, $f$, in this scaling relation and  confirm the cusp-like nature in higher dimensions.  In addition, the higher dimensional critical exponents and periods are calculated and compared to previous work.  
Our result in 7 dimensions is consistent within error to those of \cite{JB05} and inconsistent with \cite{ES05}

The paper is organized as follows. Section II reviews the formalism and derivation of the evolution equations in P-G coordinates for spherically symmetric collapse of a scalar field in arbitrary spacetime dimensions. Section III describes the numerical method and presents our results. Section IV closes with conclusions and prospects for future work.


\section{Equations of Motion}

The equations of motion are derived from the action for gravity minimally coupled to a massless scalar field,

\be
S[g_D, \psi_D] = \frac{1}{16 \pi G_D} \int{dx^D} \sqrt{(-g_D)} R_D(g_D) - \frac{1}{2} \int{dx^D} \sqrt{(-g_D)} |\nabla_D \psi_D|^2,
\label{action}
\ee

\noindent where $D$ is the number of dimensions and $G_D$, $g_D$, $R_D$ and $\psi_D$ are the $D$ dimensional gravitational constant, metric, Ricci scalar and the massless scalar field.  $|\nabla_D \psi_D|^2$ is short hand for $g_D^{\mu \nu} \partial_\mu \psi_D \partial_\nu \psi_D$.  

\subsection{Dilaton Formalism}

Since we are concerned with spherical symmetry, dimensional reduction and a field reparameterization can be used to put the action and equations in so called 2-D ``dilaton" gravity form. This simplifies the problem and allows a unified treatment of all spacetime dimensions. See \cite{DG02} for a review and \cite{GK98} for its applications to black holes. The dimensionally reduced action in dilaton form is:

\be
S[g, \phi, \psi] = \frac{1}{2} \int{dxdt \sqrt{(-g)} \left[ (1/G) (\phi R(g) + V(\phi)) -h(\phi) |\nabla \psi|^2 \right]},
\label{dilaction}
\ee

\noindent It is obtained by making the following field definitions:

\bea
G&=&\frac{n \pi}{(n-1) \nu^{(n)}} G_D, \qquad \phi = \frac{n}{8(n-1)} r^n, \qquad V(\phi) = \left( \frac{8(n-1) \phi}{n(n-1)^n} \right) ^ {-\frac{1}{n}}, \nonumber \\
h(\phi)&=&\frac{8(n-1)}{n} \phi, \qquad \psi = \sqrt{\nu^{(n)}}\psi_D, \qquad j(\phi) = \int_0^\phi d \tilde{\phi} V(\tilde{\phi}),
\label{diliton}
\eea

\noindent where $r$ is the areal radius, $G$ and $\psi$ are the spherically symmetric versions of the gravitational constant and the scalar field, $n=D-2$ is the number of angular dimensions and $\nu^{(n)}$ is the volume of a unit n-sphere. Note that the dilaton, $\phi$, has a geometrical interpretation as the area of a sphere at fixed distance from the center. With these field redefinitions the physical metric is related to the two dimensional metric, $ds^2$ as follows:

\be
ds_{physical}^2 = \frac{ds^2}{j(\phi)} + r^2(\phi)d\Omega_n^2.
\label{physicalmetric}
\ee

Using this formalism we can work with the 2D metric, $g_{\mu\nu}(r, t)$, and associated Ricci scalar, $R(r, t)$, without losing the effects of the D dimensional curvature.

\subsection{Gauge Fixing}

The derivation of a partially gauge fixed hamiltonian from the action (\ref{dilaction}) was carried out in \cite{RD07} using the general ADM metric,

\be
ds^2 := e^{2\rho}[-\sigma^2 dt^2 + (dx + Ndt)^2],
\label{ADMmetric}
\ee

\noindent where $\rho$, $\sigma$ and $N$ are arbitrary functions of the spacetime coordinates, $x$ and $t$.  The gauge chosen is given by,

\be
j(\phi)=l\phi'
\label{gaugefixr}.
\ee

This gauge choice forces the spatial coordinate to be the area radius, $r$, found in the Schwarzschild and PG metrics and gives the hamiltonian,

\be
H(\rho, \Pi_\rho, \psi, \Pi_\psi) = \int{dr} \sigma \left[-\frac{e^{2\rho}}{j(\phi)} {\cal M}' + {\cal G_M} + \frac{G\Pi_\rho\psi'\Pi_\psi}{j(\phi)} \right] + \int{dr} \left[ \frac{\sigma e^{2\rho}}{j(\phi)} {\cal M} \right]',
\label{hamiltonian}
\ee

\noindent where $\Pi_\rho$ and $\Pi_\psi$ are the conjugate momenta corresponding to $\rho$ and $\psi$.  The fields ${\cal M}$ and ${\cal G_M}$ are given by,

\be
{\cal M} = \frac{1}{2G}\left(G^2e^{-2\rho}\Pi_\rho^2-\frac{(\phi^\prime)^2}{e^{2\rho}}+j(\phi)\right),
\label{Mdef}
\ee

\be
{\cal G_M} = \frac{1}{2}\left(\frac{\Pi_\psi^2}{h(\phi)} + h(\phi)(\psi^\prime)^2\right).
\label{GMdef}
\ee

In the above ${\cal M}$ is so-called the mass function that generalizes the Misner-Sharpe mass \cite{misner_sharp} to higher dimensions. It aysmptotes to the ADM mass as $r \to \infty$.  ${\cal G_M}$ is the energy density of the free scalar field. 

We now completely fix the gauge using,

\be
e^{\rho} = \sqrt{j(\phi)},
\label{gaugefixPG}
\ee

\noindent as was done in \cite{JZ09}.  This gauge choice forces the metric into the non-static, higher dimensional generalization of the PG metric, given by

\be
ds^2 = j(\phi) \left[ -\sigma^2 dt^2 + \left( dr + \sqrt{\frac{2G {\cal M}}{j(\phi)}\sigma dt} \right)^2 \right].
\label{PG}
\ee

\noindent An advantage of the PG coordinate system is that spatial slices $t=constant$ are  regular across apparent horizons that form during evolution.  The Hamiltonian of Eq.(\ref{hamiltonian}) and gauge choices of equations (\ref{gaugefixr}) and (\ref{gaugefixPG}) give the fully reduced equations of motion \cite{RD07}, \cite{JZ09},

\be
\dot{\psi} = \sigma \left(\frac{\sqrt{2G{\cal M}} \psi^\prime}{\sqrt{j(\phi)}} + \frac{\Pi_\psi}{h(\phi)}\right),
\label{psieom}
\ee

\be
\dot{\Pi}_\psi = \left[\sigma \left( h(\phi)\psi^\prime + \frac{\sqrt{2G{\cal M}} \Pi_\psi}{\sqrt{j(\phi)}} \right) \right]^\prime,
\label{Pieom}
\ee

\be
{\cal M}^\prime =  {\cal G_M}+\psi^\prime \Pi_\psi \sqrt{\frac{2G{\cal M}}{j(\phi)}},
\label{Meom}
\ee

\be
\sigma^\prime + \frac{G \psi^\prime\Pi_\psi }{\sqrt{2G{\cal M} j(\phi)} } \sigma = 0.
\label{sigmaeom}
\ee

Notice that these equations depend on the number of dimensions via Eq.(\ref{diliton}).  

\section{Methodology and Results}

\subsection{Computational Details}

In order to solve the equations of motion we used the initial conditions:  

\begin{eqnarray}
\psi &=& A r^2 exp \left[ -\left( \frac{r-r_0}{B} \right)^2 \right], \nonumber \\
\Pi_\psi &=& 0,
\label{eq:initialpsi}
\end{eqnarray}

\noindent where $A$, $B$ and $r_0$ are parameters that can in principle be varied to study mass scaling. These initial conditions are the same as those used in \cite{JZ09}. We verified universality of our results by varying the amplitude, $A$, and the width, $B$, of the initial pulse. 

The equations of motion, (\ref{psieom})
-(\ref{sigmaeom}), were solved using a forth order Runge-Kutta scheme with deriviatives being calculated using the finite difference method.

In order to maintain stability we used an adaptive time step, $\Delta t(t)$:
\be
\Delta t(t) = min_r \left[ \frac{dt}{dr} \Delta r(r) \right],
\label{timespacing}
\ee

\noindent where $\Delta r(r)$ is the spacing of the spatial lattice and $\frac{dt}{dr}$ is the inverse of the local speed of an ingoing null geodesic.

Much of the interesting behaviour of the collapse near criticality occurs near the origin.  This requires close spacing of the spatial mesh near $r=0$.  It is not computationally realistic, however, to use this spacing along the entire spatial slice, which needs to be long enough so that none of the mass leaves during the simulation. The use of close spacing for the entire slice would dramatically increase the simulation time.  For this reason the spacing near the origin was set to $10^{-5}$ and then smoothly increased to $10^{-2}$ over the first 100 of the 1200 total lattice points.

\subsection{Results}

To find Choptuik's mass scaling relation we first found the critical values, $A_*$ and $B_*$, using a bisection method to an accuracy of $\Delta A_*/A_*  \approx \Delta B_*/B_* \approx 10^{-15}$.  We then varied the parameters $A$ and $B$ in our initial data function, Eq.(\ref{eq:initialpsi}), and found the mass of the black hole at formation, $M_{BH}$, as a function of the initial parameters.  An apparent horizon forms when an outgoing null geodesic becomes momentarily stationary. For a metric in P-G form as in Eq.(\ref{PG}), this is signalled by the condition:
\be
\left. 2G{\cal M} - j(\phi)\right|_{r_f} =0,
\ee
where $r_f$ is the areal radius of the horizon on formation. The mass of the black hole at horizon formation is then given by the value of the mass function ${\cal M}(r_f)$ at that point. We emphasize again that this is not the ADM mass of the final static black hole that is left behind once all the matter has fallen through the horizon. 

\begin{figure}[ht!]
\centering
\subfigure[4 Dimensions]{
\includegraphics[width=0.4\linewidth]{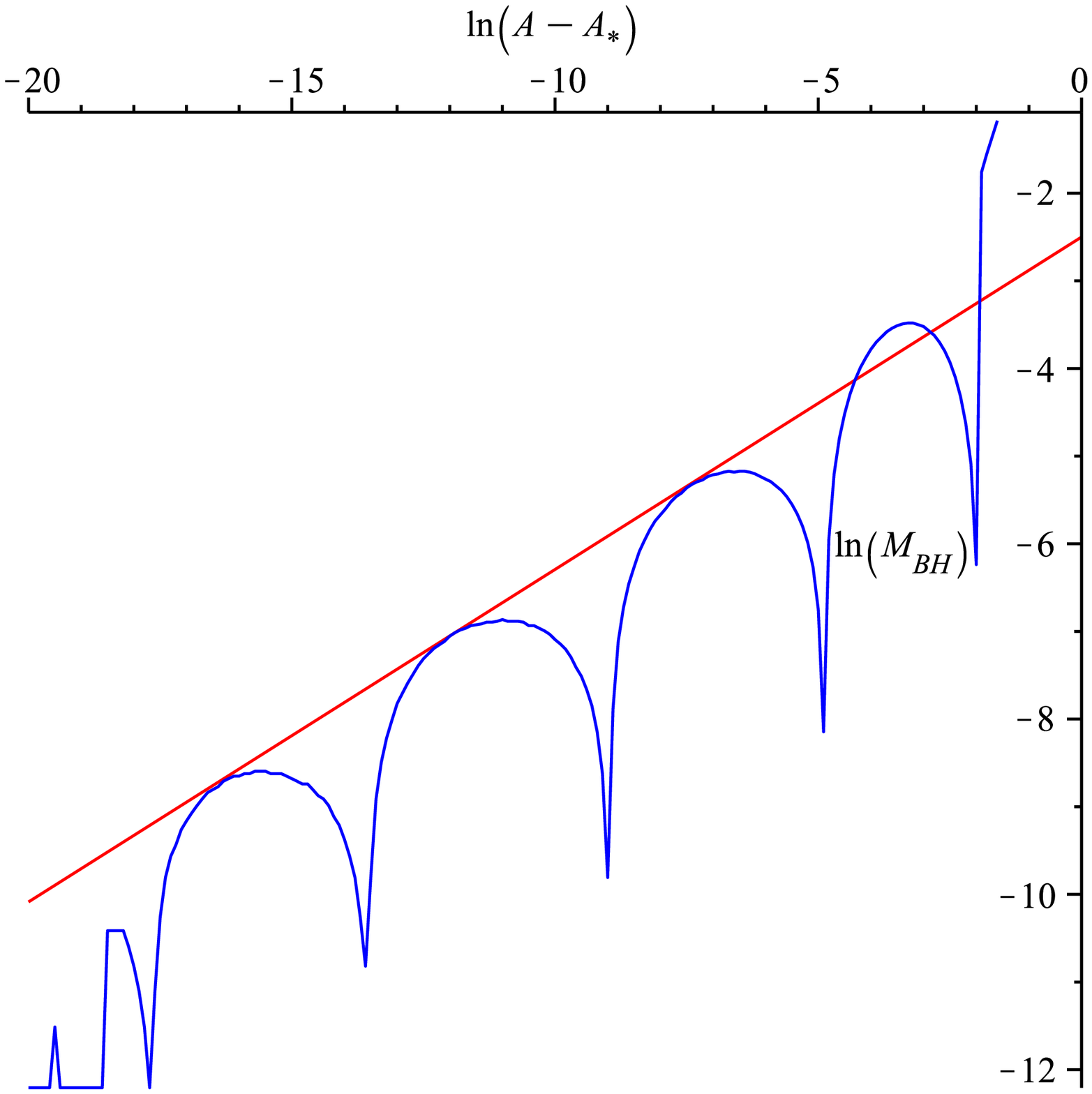}
\label{4DChop}
}
\hspace{0.25in}
\subfigure[5 Dimensions]{
\includegraphics[width=0.4\linewidth]{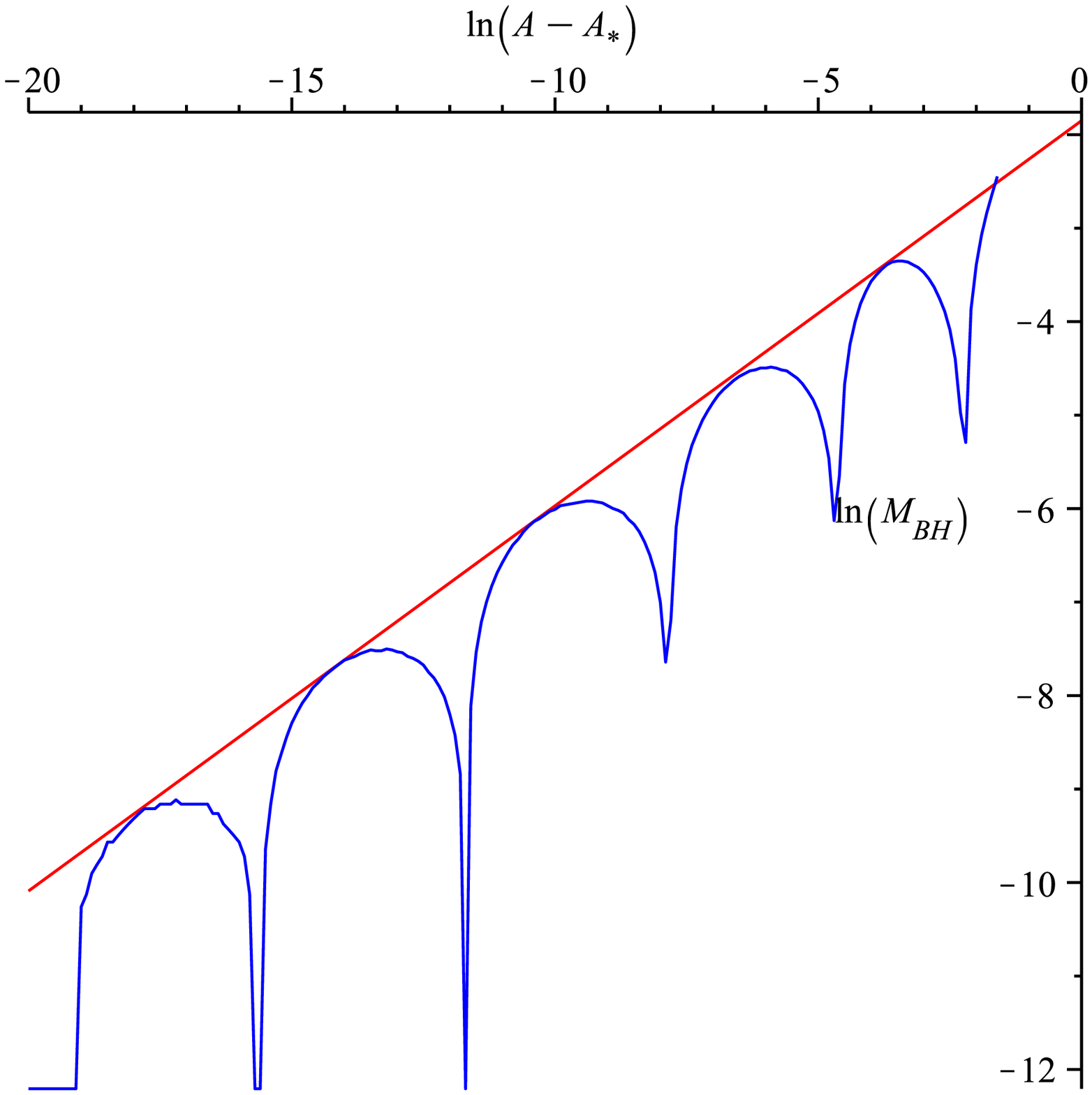}
\label{5DChop}
}
\subfigure[6 Dimensions]{
\includegraphics[width=0.4\linewidth]{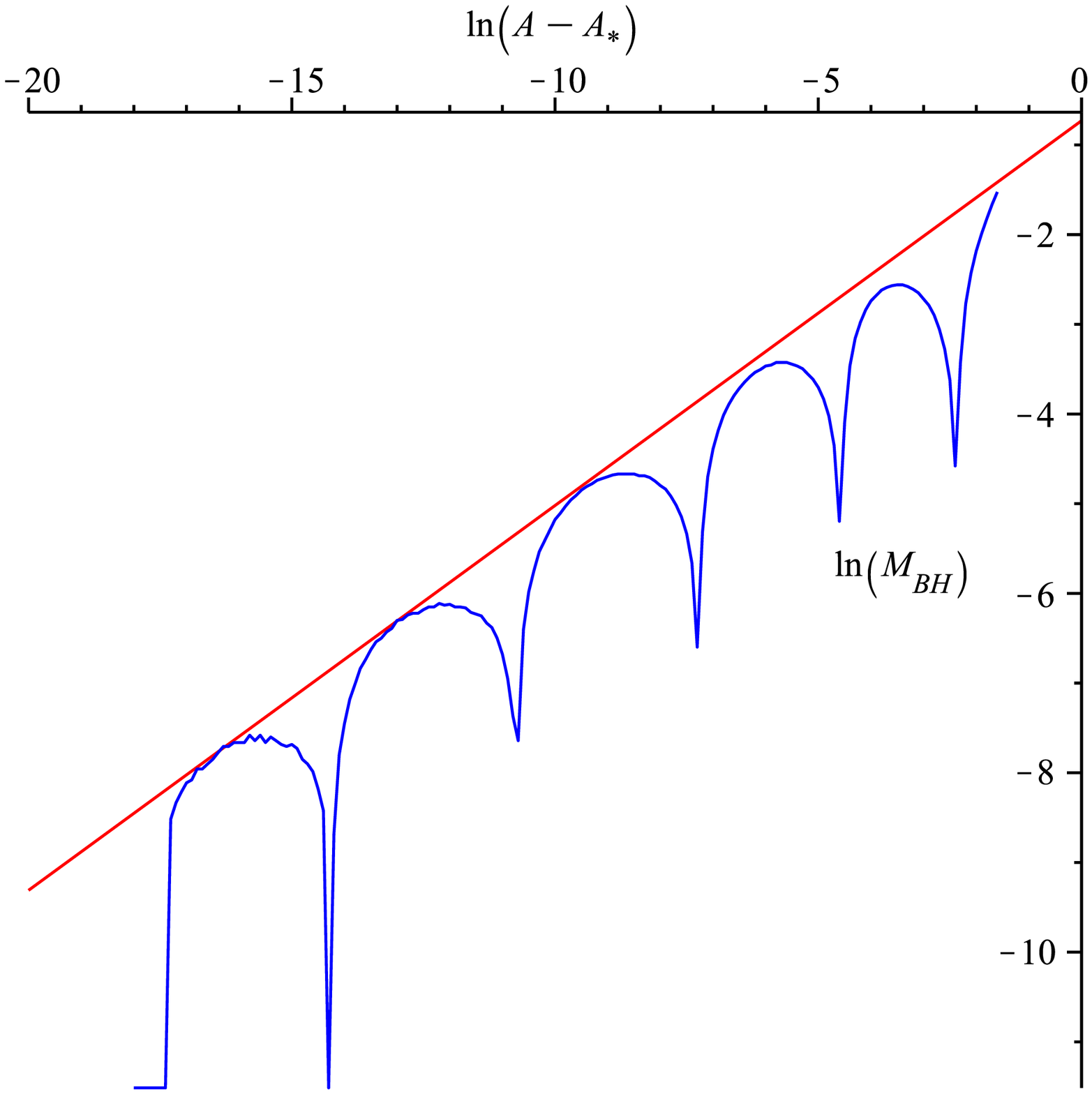}
\label{6DChop}
}
\hspace{0.25in}
\subfigure[7 Dimensions]{
\includegraphics[width=0.4\linewidth]{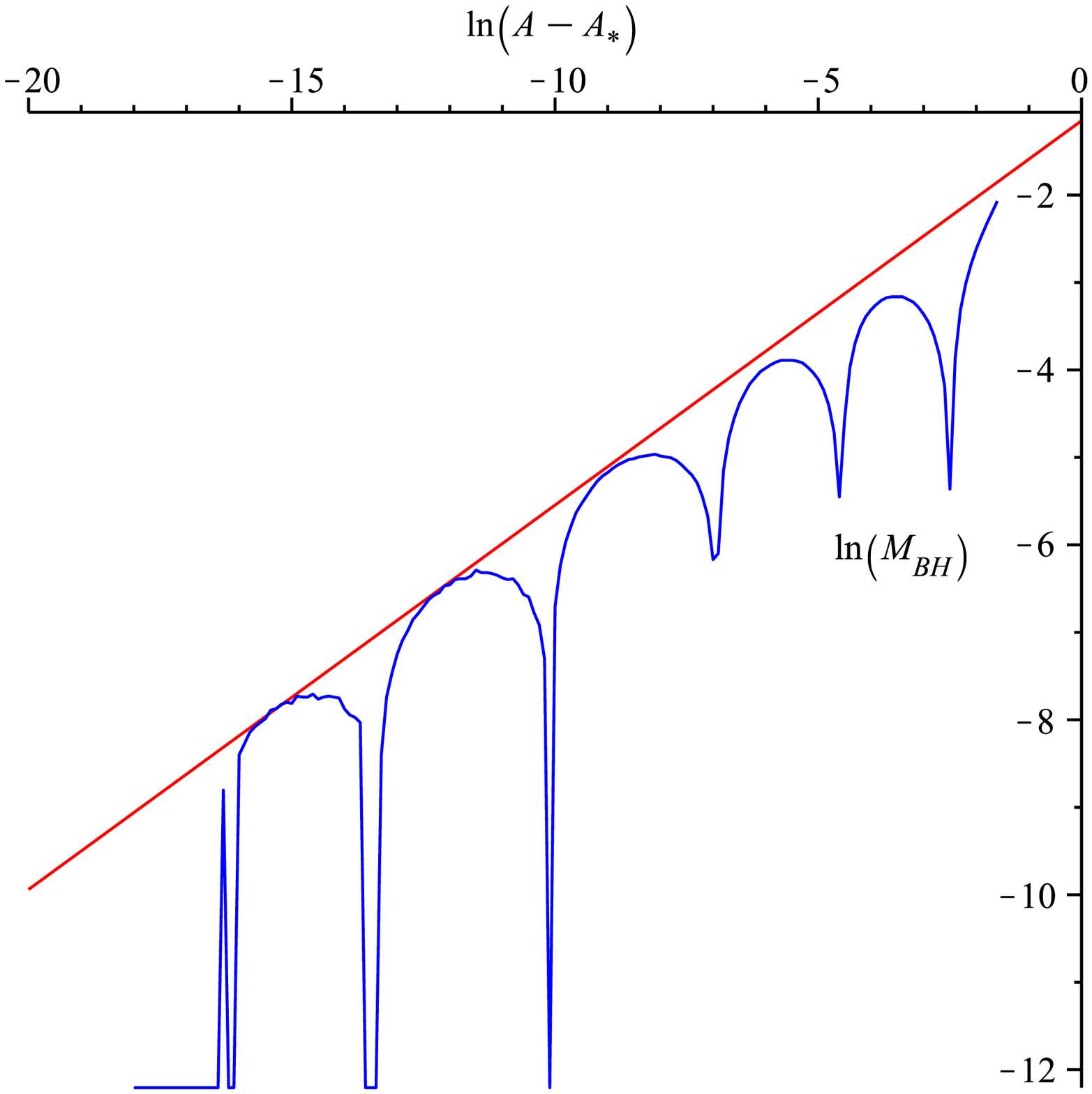}
\label{7DChop}
}
\caption[Mass scaling of A parameter]{Mass Scaling in 4, 5, 6 and 7 dimensions with the $A$ parameter varied}
\label{mass scaling plots}
\end{figure}

The mass scaling plots can be seen in Fig.(\ref{mass scaling plots}).  A straight line, which osculates the curve, has been plotted with the data to illustrate the linear term in the mass scaling relation, Eq.(\ref{Chop}).  Notice that these plots confirm the existence in 5, 6 and 7 dimensions of the cusps in the periodic function  that were originally noted in four dimensions in \cite{JZ09}.  It is also important to note from Fig.(\ref{mass scaling plots}) that  as we move away from the near critical region (i.e. as $A-A_*$ gets large) the cusps systematically move away from the critical straight line.  In 4 dimensions the results move above the critical line, in five dimensions they more or less stay on the straight line, but in higher dimensions the cusps move downward. This illustrates graphically that in higher dimensions the critical exponent will be underestimated if one is not sufficiently close to criticality.  

Points in the data that were close to the straight lines ($< 2 \%$) of Fig.(\ref{mass scaling plots}) were chosen and were used to find the critical exponent using linear regression.  An estimate of error was found by slightly varying which points were chosen to calculate the slope.  The period of $f$ in Eq.( \ref{Chop}) was calculated by measuring the distance between the cusps of the plots.  A measure of error for these cusps was taken as the largest deviation from the average.  All measurements were made using the first three periods of the data where the near criticality approximation is valid.  This analysis was done for both of the cases where $A$ and $B$ were varied in the initial data.  The results can be seen in table (\ref{gammadelta}).  These results show good agreement with the results of Bland et. al. \cite{JB05} and Ziprick and Kunstatter \cite{JZ09}.  

\begin{figure}[ht!]
\centering
\includegraphics[width=0.4\linewidth]{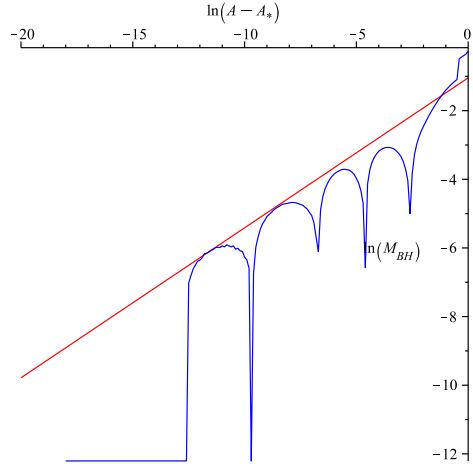}
\caption[8D Mass scaling of A parameter]{Mass Scaling in 8 dimensions ($A$ varied)}
\label{8DChop}
\end{figure}

In order to give a clearer sense of why our results are limited to seven dimensions, we present eight dimensional data in Fig.(\ref{8DChop}).  In lower dimensions we were able to obtain supercritical data down to  $ln(A-A_*) \approx -17$. In eight dimensions we were limited to $ln(A-A_*)<-12.6$ without decreasing the lattice spacing near the origin, which in turn would dramatically increase the simulation time.  The effect of this is seen in figure (\ref{8DChop}):  we are unable to get close enough to criticality for the slope to be approximately constant over three periods.  For this reason we were unable to reliably calculate the critical exponent for eight dimensions with the same accuracy as lower dimensional simulations.  If we nonetheless obtain a slope from  the first two periods of the simulation (as opposed to three) then we find $\gamma = 0.44 \pm 0.02$ and $T = 3.0 \pm 0.1$.  This result lies between that of Bland et. al. \cite{JB05} ($\gamma = 0.4459 \pm 0.0054$, $T = 3.11 \pm 0.1$) and Sorkin and Oren \cite{ES05} ($\gamma = 0.436 \pm 2\%$, $T = 3.1 \pm 3\%$).  It should be noticed that given the arguments above and the negative curvature that we see in figure (\ref{8DChop}) our current value is almost certainly an underestimate. That is, a third cusp closer to criticality would most likely raise the value of the slope, bringing the critical exponent closer to that of \cite{JB05}. 

\begin{table}[ht]
\caption{Critical Exponent, $\gamma$ and Period, $T$ for Four to Seven Dimensions, $D$}
\centering
\begin{tabular}{c c c c}
\hline\hline 
D &  & $\gamma$ & T \\ 
\hline
4 & A varied & 0.378 $\pm$ 0.002 & 4.4 $\pm$ 0.3 \\
  & B varied & 0.379 $\pm$ 0.002 & 4.4 $\pm$ 0.2 \\
  & from \cite{JB05} & 0.374 $\pm$ 0.002 & 4.55 $\pm$ 0.1 \\
  & from \cite{JZ09} & 0.375 $\pm$ 0.004 & 4.6 $\pm$ 0.1 \\
  & from \cite{ES05} & 0.372 $\pm$ 1\% & 4.53 $\pm$ 2\% \\
\hline
5 & A varied & 0.413 $\pm$ 0.002 & 3.9 $\pm$ 0.8 \\
  & B varied & 0.416 $\pm$ 0.002 & 3.7 $\pm$ 0.4 \\
  & from \cite{JB05} & 0.412 $\pm$ 0.004 & 3.76 $\pm$ 0.1 \\
  & from \cite{ES05} & 0.408 $\pm$ 2\% & 4.29 $\pm$ 2\% \\
\hline
6 & A varied & 0.429 $\pm$ 0.003 & 3.4 $\pm$ 0.3 \\
  & B varied & 0.428 $\pm$ 0.002 & 3.3 $\pm$ 0.2 \\
  & from \cite{CCD99}&0.424 &3.03\\
  & from \cite{JB05} & 0.430 $\pm$ 0.003 & 3.47 $\pm$ 0.1 \\
  & from \cite{ES05} & 0.422 $\pm$ 2\% & 4.05 $\pm$ 2\% \\
\hline
7 & A varied & 0.440 $\pm$ 0.005 & 3.1 $\pm$ 0.1 \\ 
  & B varied & 0.440 $\pm$ 0.006 & 3.1 $\pm$ 0.4 \\
  & from \cite{JB05} & 0.441 $\pm$ 0.004 & 3.36 $\pm$ 0.1 \\
  & from \cite{ES05} & 0.429 $\pm$ 2\% & 3.80 $\pm$ 2\% \\
\hline\hline
\end{tabular}
\label{gammadelta}
\end{table}

\section{Conclusion}

We have successfully confirmed the existence of cusps in the mass scaling function, Eq.(\ref{Chop}), in 4 to 8 dimensions as observed in 4 dimensions by Ziprick and Kunstatter \cite{JZ09}.  In addition the mass scaling plots obtained using PG coordinates give critical data which agrees with previous results in 4, 5 and 6 dimensions.  However, in 7 dimensions, our critical exponent and echoing period agree with Bland et. al. \cite{JB05} but disagree with the values obtained by Sorkin and Oren \cite{ES05}. It is important to note that \cite{JB05} claimed that the critical exponent was a monotonic function of spacetime dimension that asympotes to 1/2, whereas \cite{ES05} suggested that the critical exponent peaks near D=10. While both claims are intriguing, only one (at most) can be right. Since we are limited at the moment to $D\leq7$ we cannot make any definitive claims about the asympotic behaviour. Our result in $D=7$ supports that of \cite{JBthesis}, as does our more tentative result in $D=8$. It is however impossible to make definitive claims about the asymptotic behaviour without pushing the PG calculation to higher dimensions. 


Another next step in our work will be to incorporate loop quantum gravity inspired corrections which have been shown to resolve the singularity \cite{JZ09_2} in four dimensions.  It will be interesting to check the dependence of singularity resolution on dimensionality as well as verify the existence of a mass gap for higher dimensional, quantum corrected mass scaling.

\section{Acknowledgements}

We thank Jonathan Ziprick for providing many helpful discussions on physics and code. We are extremely grateful for the help and encouragement of Randy Kobes during the early stages of this work. Dr. Kobes passed away tragically in September, 2010. We also thank Mathias Pielahn, Ari Peltola and Alex Neilsen for insightful conversations. Tim Taves would like to thank Jonathan Wheelwright for computational skills and resources and the University of Manitoba - Faculty of Science for funding. The work was supported in part by the Natural Science and Engineering Research Council of Canada.

\end{document}